\documentstyle[psfig,12pt,a4wide]{article}
\def\eqn{\begin{equation}} 
\def\eeqn{\end{equation}}
\def\arr{\begin{array}} 
\def\earr{\end{array}}
\def\eqna{\begin{eqnarray}} 
\def\eeqna{\end{eqnarray}} 
\def\a{\alpha}
\def\b{\beta} 
\def\D{\Delta}
\def\s{\sigma} 
\def\d{\delta}
 
\def\o{\omega}

\def\e{\epsilon}
 
\def\m{\mu} 
\def\n{\nu} 
\def\la{\lambda}

\def\p{\partial} 
 
\font\mybb=msbm10 at 12pt 
\def\bb#1{\hbox{\mybb#1}}

\begin{document}

\vspace*{-.6in}
\thispagestyle{empty}
\begin{flushright}
PUPT-1926\\
\end{flushright}

\vspace{.3in}
{\Large 
\begin{center}
{\bf A Test of the AdS/CFT Duality on the \\ Coulomb Branch}
\end{center}}
\vspace{.3in}
\begin{center}
Miguel S. Costa\footnote{miguel@feynman.princeton.edu}\\
\vspace{.1in}
\emph{Joseph Henry Laboratories\\ Princeton University \\ 
Princeton, New Jersey 08544, USA}
\end{center}

\vspace{.5in}

\begin{abstract}
We consider the ${\cal N}=4$ $SU(N)$ Super Yang Mills theory on the 
Coulomb branch with gauge symmetry broken to $S(U(N_1)\times U(N_2))$.
By integrating the W particles, the effective action near the 
IR $SU(N_i)$ conformal fixed points is seen to be a deformation 
of the Super Yang 
Mills theory by a non-renormalized, irrelevant, dimension 8 operator. 
The correction to the two-point
function of the dilaton field dual operator near the IR
is related to a three-point 
function of chiral primary operators at the conformal fixed points
and agrees with the classical gravity prediction, including the 
numerical factor.
\end{abstract}

\newpage

\section{Introduction}

Recently there has been a considerable effort in studying the 
AdS/CFT duality \cite{Mald,GKP,Witten} when conformal invariance 
is broken (see \cite{review} and references there in). 
The general philosophy is that conformal
invariance can be broken by considering a non-conformal vacuum of the conformal
theory, or otherwise by considering a deformation of its Lagrangian. In both
cases the dual gravitational background will have an AdS boundary but it will 
differ from this space in the inside. Identifying the radial coordinate with 
the field theory energy scale the geometry will describe the RG flow 
starting from the UV conformal theory at the AdS boundary.

In this note we shall consider the case of a non-conformal vacuum of 
$SU(N)$ ${\cal N}=4$ Super Yang-Mills (SYM) theory. 
The moduli space of this theory is given by
$\bb{R}^{6(N-1)}/S_N$, parameterizing the relative position of $N$ identical 
D3-branes in the transverse space $\bb{E}^6$. Separating the D3-branes in 
two stacks of $N_1$ and $N_2$ branes by a distance $2\D$ is equivalent to 
Higgs the theory by giving an expectation value to the scalar fields 
according to
\eqn
\frac{\langle\vec{Y}\rangle}{2\pi\a'}\equiv 
\langle \vec{\phi} \rangle =\frac{\vec{\D}}{2\pi\a'}
\left(
\arr{cc}
I_1 & 0 \\
0 & -I_2
\earr
\right)\ ,
\label{Higgs}
\eeqn
where $I_i$ is the $N_i\times N_i$ unit matrix. The gauge symmetry is broken
to $S(U(N_1)\times U(N_2))$ and the theory is left with 16 conserved 
supercharges because conformal symmetry is broken. The symmetry breaking 
process gives rise to massive W particles with mass $m_W=\D/(\pi\a')$. 
This construction provides an example of maximally
supersymmetric RG flow: starting from the UV $SU(N)$ conformal 
theory we flow to the IR conformal fixed points with $SU(N_i)$ gauge 
symmetry. Related aspects of the AdS/CFT duality can be found in references
[5-23].

To derive the dual gravitational background we start with the double-centered
D3-brane geometry with metric
\eqn
ds^2=H^{-1/2}ds^2(\bb{M}^4)+H^{1/2}ds^2(\bb{E}^6)\ ,
\label{D3-metric}
\eeqn
where the harmonic function $H$ reads
\eqn
H=1+\frac{R_1^{\ 4}}{|\vec{y}-\vec{\D}|^4}
+\frac{R_2^{\ 4}}{|\vec{y}+\vec{\D}|^4}\ ,
\label{harmf}
\eeqn
with $R_i^{\ 4}=4\pi g_s\a'^2N_i$ and $\vec{y}$ the coordinates
in the transverse space $\bb{E}^6$. Next, we take the decoupling
limit of \cite{Mald} which corresponds to drop the factor 1 
in the harmonic function
$H$ and to consider energy scales in the brane theory much smaller than 
the string scale, i.e. $\e,m_W\ll 1/\sqrt{\a'}$. We 
further require the string coupling $g_s$ to be small and $g_sN_i$ to be 
large for the supergravity approximation to hold.

In a recent paper \cite{Costa} we studied the classical 
absorption for a minimally
coupled scalar, e.g. the dilaton field, in the double-centered D3-brane
geometry. The analysis was valid for low energies such that 
$\e\ll m_g\ll m_W$, where $m_g=m_W/\sqrt{gN}$ is the gravity mass gap. In
the above decoupling limit the cross-section  for absorption by the 
$i$-th hole was seen to be
\eqn
\s_i=\frac{\pi^4\o^3R_i^{\ 8}}{8}
\left[ 1-\frac{(\o R_i)^4}{12}\left(\frac{R_j}{2\D}\right)^4
\log{\left(\frac{\o R_i^{\ 2}}{\D}\right)}+\cdots\right]\ ,
\ \ \ \ (j\ne i)\ .
\label{csection}
\eeqn
In the field theory side, the total absorption cross section 
(i.e. $\s=\s_1+\s_2$) for a scalar
field $\phi$ is related to the two-point function of the field theory dual
operator ${\cal O}_{\phi}$
\eqn
\Pi(x)=\left\langle {\cal O}_{\phi}^{\phantom{1}}(x)
{\cal O}_{\phi}(0)\right\rangle\ ,
\label{2pf}
\eeqn
calculated in the non-conformal vacuum described above. In the case of the
dilaton field the exact relation is \cite{GubserKleb}
\eqn
\s=\frac{2\kappa^2}{2i\o}\left.{\rm Disc}^{\phantom{1}}
\Pi(s)\right|_{{p^0 = \omega  \atop \vec{p} =\vec{0}}}\ ,
\label{disc}
\eeqn
where $\kappa$ is the gravitational coupling, $s=-p^2$ and $\Pi(s)$ 
is the Fourier transform of $\Pi(x)$. 

The low energy expansion of the classical cross section (\ref{csection})
corresponds to consider the dual field theory near the $SU(N_i)$ IR fixed 
point. The purpose of this letter is to determine the exact form of the 
effective action near the IR fixed points that results from integrating out 
the W particles and therefore to reproduce exactly the logarithmic correction
to the cross section using the dual field theory. We shall see that the 
effective action is given by a deformation of the ${\cal N}=4$ SYM theory 
by a non-renormalized, irrelevant, dimension 8 operator as anticipated 
in \cite{Intr,Costa}. Then, the correction to the cross section 
will be related to a three-point function of chiral
primary operators calculated at the IR fixed points \cite{GHKK}. 
To obtain exact 
agreement with the classical cross section calculation it was essential to
use the symmetrized trace of the Yang Mills fields  in the deformed
Lagrangian and to keep only the planar diagrams. This result is further
evidence for a non-renormalization theorem for three-point functions of chiral
primary operators [27-32] and provides a test of the 
AdS/CFT duality on the Coulomb branch.

\section{Effective Action in the Infrared}

We start by writing the Lagrangian for the bosonic sector of 
${\cal N}=4$ $SU(N)$ SYM theory in the following form
\eqn
{\cal L}_0=-\frac{1}{4}{\rm tr}\left[ F_{AB}F^{AB} \right]\ ,
\label{Lagrangian}
\eeqn
where $A,B,\cdots$ are ten-dimensional indices and $F_{AB}$ is short for 
$F_{\m\n}=D_{\m}A_{\n}-D_{\n}A_{\m}$, $F_{\m m}=D_{\m}\phi_m$ with  
$D_{\m}\equiv \p_{\m}+ig_{YM}[A_{\m},\ ]$ and $F_{mn}=ig_{YM}[\phi_m,\phi_n]$.
We removed the gauge coupling from the front of the action by rescaling the
fields as $(A_{\m},\phi^m)\rightarrow g_{YM} (A_{\m},\phi^m)$, which also
rescales the expectation value for the scalars in equation (\ref{Higgs}). 

We want to integrate the W's in order to obtain an effective action for the
light $SU(N_1)$ and $SU(N_2)$ coloured fields at energy scales $\e\ll m_W$
as explained in \cite{Costa}.
This is similar to the probe calculations extensively considered in 
the literature [33-36], where $N_1=N$ and $N_2=1$ . 
In the latter case the first 
non-vanishing one-loop diagram involves 4 $SU(N)$ coloured legs, with the 
following contribution to the effective bosonic
Lagrangian \cite{Mald2,ChepTsey}
\eqn
{\cal L}_1=-\frac{\pi^2g^4_{YM}}{(2\D/\a')^4}
{\rm Str}\left[F_{AB}F^{BC}F_{CD}F^{DA}-
\frac{1}{4}\left(F_{AB}F^{AB}\right)^2\right]\ .
\label{1loop}
\eeqn
Notice that we are using the symmetrized trace as argued in 
\cite{ChepTsey}. This fact will be essential to obtain agreement with the
dual classical calculation of the absorption cross-section. 
Also, these $F^4$ terms are protected \cite{DineSeiberg}, 
i.e. they are not renormalized by higher loop diagrams and therefore 
comparison with the strongly coupled supergravity regime is allowed.

Now we generalize the above result to arbitrary large $N_1$ and $N_2$. The
only difference is that the term in the effective action involving 4
$SU(N_i)$ fields will be multiplied by a factor $N_j$ $(i\ne j)$ 
because we have a
$SU(N_j)$ colour index to sum over around the loop (due to the W's exchange).
Also, for large $N$ graphs with both $SU(N_1)$ and $SU(N_2)$ coloured 
external legs are subleading since they are associated with non-planar 
graphs. Hence, in the IR and for large $N$ there are no terms in the 
effective action of the type ${\rm tr_1}(F^2){\rm tr_2}(F^2)$, where
the subscript in ${\rm tr}_i$ means that the trace is over $SU(N_i)$ fields.
We conclude that the $F^4$ terms in the effective action for the IR 
$SU(N_i)$ theory read
\eqn
{\cal L}_1=-\frac{\pi^2g^4_{YM}}{(2\D/\a')^4}
N_j{\rm Str}_i\left[F_{AB}F^{BC}F_{CD}F^{DA}-
\frac{1}{4}\left(F_{AB}F^{AB}\right)^2\right]\ ,\ \ \ \ (j\ne i)\ .
\label{effaction}
\eeqn

To check this result consider the action for a probe of $N_i$ D3-branes 
in the AdS near-horizon geometry of $N_j$ D3-branes, i.e. we assume that 
$N_j\gg N_i$. The probe dynamics is determined by the non-abelian DBI 
action \cite{Tsey} which describes the effect of integrating all 
the massive strings stretching between the probe and the branes. 
The AdS background is described
by the metric (\ref{D3-metric}) with the harmonic function 
$H\equiv f=(R_j/r)^4$. If the center 
of mass for the $N_i$ probes is placed at $r$ we have 
\eqn
S_{probe}=-T_3\int d^4x f^{-1} {\rm Str}_i
\left[\sqrt{-{\rm det}\left(
\eta_{\a\b}+f\p_{\a}Y^m\p_{\b}Y_m+2\pi\a'\sqrt{f}F_{\a\b}
\right)}-I_i\right]\ ,
\label{probe}
\eeqn
where $T_3=((2\pi)^3g_s\a'^2)^{-1}$ is the D3-brane tension.
Next we give an expectation value to the scalars 
$\langle \vec{Y}\rangle=2\vec{\D}I_i$, which means that $r=2\D+\d r$ where
$\d r$ is the center of mass fluctuating field in the radial direction.
For large $N_i$ we can discard the fields in the center of the gauge 
group and consider only $SU(N_i)$ fields. Thus, setting 
$f=(R_j/2\D)^4$, expanding the DBI action and rescaling the fields 
according to 
$(A_{\a},\phi_m=(2\pi \a')^{-1}Y_m)\rightarrow g_{YM}(A_{\a},\phi_m)$
we obtain the probe Lagrangian
\eqn
{\cal L}_{probe}=
-\frac{1}{4}{\rm tr}_i\left[F_{AB}F^{AB}\right]
-\frac{\pi^2g^4_{YM}N_j}{(2\D/\a')^4}
{\rm Str}_i\left[F_{AB}F^{BC}F_{CD}F^{DA}-
\frac{1}{4}\left(F_{AB}F^{AB}\right)^2\right]\ ,
\label{probe1}
\eeqn
which is in agreement with the previous result as expected from the
non-renormalization of the $F^4$ terms.

In the context of the asymptotically flat D3-brane geometry, the form 
of the above deformation of the SYM Lagrangian was conjectured based
on the DBI corrections to the SYM theory \cite{GHKK}, or alternatively on the
basis of $PSU(2|2,4)$ representation theory \cite{GubserHash}. For a 
geometry with harmonic function
\eqn
H=h+\left(\frac{R}{r}\right)^4\ ,
\label{harmonic}
\eeqn
the dual field theory was conjectured to be \cite{GubserHash,Intr}
\eqn
{\cal L}={\cal L}_0-\frac{h}{8T_3}{\cal O}_8\ ,
\label{dualLagr}
\eeqn
where ${\cal O}_8$ is an irrelevant, dimension 8 operator. This
deformation is irrelevant in the IR which agrees with the fact that for
$r\rightarrow 0$ the constant $h$ becomes irrelevant in the 
harmonic function $H$.
What remains an open question is if the Lagrangian (\ref{dualLagr}) 
describes the dual gravity theory as we flow from the IR. If we
assume that the DBI action is dual to the full D3-brane geometry
$(h=1)$, then we can regard (\ref{dualLagr}) as the first correction
to the SYM theory and determine ${\cal O}_8$ to be exactly
${\cal O}_8={\rm Str}[F^4-\frac{1}{4}(F^2)^2]$. However, even in this
case agreement between the gravity and the field theory calculations 
of the cross section is not found \cite{GHKK}. 
Fortunately the case here studied is entirely under control. 
While in the asymptotic flat space case the DBI action
arises from integrating the massive string states that are dropped
out in the decoupling limit, in our case the deformation of the IR 
Lagrangian arises from integrating the W's that do survive the decoupling
limit. In fact, in the decoupling limit of the double-centered
D3-brane geometry the harmonic function $H$ is given down the $i$-th 
throat by
\eqn
H=\left(\frac{R_j}{2\D}\right)^4+\left(\frac{R_i}{r}\right)^4\ .
\label{harmonic2}
\eeqn
Then the deformation of the SYM action in the IR $SU(N_i)$ conformal 
fixed point is indeed given by (\ref{dualLagr}) with $h=(R_j/2\D)^4$
and ${\cal O}_8={\rm Str}_i[F^4-\frac{1}{4}(F^2)^2]$ \cite{Intr,Costa}. 
This (non-renormalized) deformation
was computed exactly as a result of integrating the W's. Hence, if we
believe the AdS/CFT correspondence to hold on the Coulomb branch, the 
gravity and perturbative field theory calculation of protected
quantities in the IR using the Lagrangian (\ref{dualLagr}) must
give exactly the same answer.

\section{Field Theory Calculation of Cross-Section}

We proceed by calculating the cross section for absorption of the dilaton
field using the field theory approach. This calculation was done in
\cite{GHKK} using a $U(1)$ model. We refer the reader to that reference
for the details and will present here only the relevant steps necessary
to obtain the correct answer. For world-volume on-shell processes
that involve the coupling of the dilaton to the brane only the gauge
field is relevant. In the IR $SU(N_i)$ theory the operator 
${\cal O}_{\phi}$ dual to the dilaton field reads
\eqn
\arr{rcl}
S_{int}&=&
\displaystyle{\int d^4x\ \phi\ \frac{1}{4}
\left( {\rm tr}_i\left[F_{\m\n}F^{\m\n}\right]
+\frac{h}{T_3}{\rm Str}_i\left[F_{\m\n}F^{\n\eta}F_{\eta\la}F^{\la\m}-
\frac{1}{4}\left(F_{\m\n}F^{\m\n}\right)^2\right]\right)}\ ,
\phantom{\rule[-7mm]{0.15mm}{5mm}}\\
&\equiv&
\displaystyle{\int d^4x\ \phi\ {\cal O}_{\phi}
\equiv \int d^4x\ \phi\frac{1}{4}
\left[{\cal O}_4+\frac{h}{T_3}{\cal O}_8\right]}\ .
\earr
\label{inter}
\eeqn
Then the two-point function for the operator 
${\cal O}_{\phi}$ is
\eqn
\arr{rcl}
\Pi(x)&=&\displaystyle{
\left\langle {\cal O}_{\phi}^{\phantom{1}}(x){\cal O}_{\phi}(0)
\right\rangle_{h=\left(R_j/2\D\right)^4}}
=\int{\cal D}A_{\mu}
e^{-\int d^4z\left[\frac{1}{4}{\cal O}_4+\frac{h}{8T_3}{\cal O}_8\right]}
{\cal O}_{\phi}(x){\cal O}_{\phi}(0)
\phantom{\rule[-7mm]{0.15mm}{5mm}}\\
&=&
\displaystyle{\left\langle {\cal O}_{\phi}(x){\cal O}_{\phi}(0) 
\left( 1-\left(\frac{R_j}{2\D}\right)^4\frac{1}{8T_3}
\int d^4z{\cal O}_8(z) \right)\right\rangle_{h=0}}
\phantom{\rule[-8mm]{0.15mm}{5mm}}\\
&=&
\displaystyle{
\frac{1}{2^4}
\left\langle {\cal O}_4^{\phantom{1}}(x){\cal O}_4(0)\right\rangle_{h=0}
-\left(\frac{R_j}{2\D}\right)^4\frac{1}{2^7T_3}
\int d^4z\left\langle{\cal O}_4^{\phantom{1}}(x){\cal O}_8(z)
{\cal O}_4(0)\right\rangle_{h=0}}
\phantom{\rule[-6mm]{0.15mm}{5mm}}\\
&\equiv&
\Pi_0(x)+\Pi_1(x)\ ,
\earr
\label{2ptf}
\eeqn
where we are just keeping the leading terms that will give the logarithmic 
corrections in the classical gravity result for the cross section. 
We see that the correction to the two-point function is related to a 
three-point function of chiral primary operators calculated
at a IR conformal fixed point.

We start by writing the Euclidean space propagator for the field 
strength $(F_{\m\n})^{ab}$ in the conformal theory
\eqn
\arr{c}
\displaystyle{(
F
\rule[4mm]{0.15mm}{1.65mm}
\overline{^{\phantom{\frac{1}{1}}}_{\m\n})^{ab}(x)(F}
\rule[4mm]{0.15mm}{1.65mm}
_{\a\b})^{cd}(0)
\equiv
\left\langle
\left(F_{\m\n}\right)^{ab}(x)\left(F_{\a\b}\right)^{cd}(0)\right\rangle=}\\
\displaystyle{=\frac{\d^{ad}\d^{bc}}{\pi^2x^4}
\left[\d_{\m\a}\d_{\n\b}-\d_{\n\a}\d_{\m\b}
-\frac{2}{x^2_{\phantom{\frac{1}{1}}}}
\left(\d_{\n\b}^{\phantom{1}}x_{\m}x_{\a}+\d_{\m\a}x_{\n}x_{\b}
-\d_{\n\a}x_{\m}x_{\b}-\d_{\m\b}x_{\n}x_{\a}\right)\right]}\ ,
\earr
\label{propagator}
\eeqn
where $a,b,\cdots$ are $SU(N_i)$
colour indices. In what follows it is convenient 
to write first the following contraction of the field strength propagators
\eqn
\left( \phantom{\rule[4mm]{0.05mm}{2.8mm}}\right. 
(F
\rule[4.6mm]{0.15mm}{2.77mm}
\overline{\phantom{\rule[4mm]{0.05mm}{2.8mm}}
_{\a\b})^{ab} (F
\rule[4.6mm]{0.15mm}{1.65mm}
\overline{^{\a\b})^{ba}
\left.\phantom{\rule[4mm]{0.05mm}{1.65mm}}\right)
(x)\left(\phantom{\rule[4mm]{0.05mm}{1.65mm}}\right. (F}
\rule[4.6mm]{0.15mm}{1.65mm}
_{\m\n})^{cd} (F}
\rule[4.6mm]{0.15mm}{2.77mm}
_{\eta\la})^{ef}
\left.\phantom{\rule[4mm]{0.05mm}{1.65mm}}\right)(0)
=
\frac{2\d^{de}\d^{cf}}{\pi^4x^8}
\left(\d_{\m\eta}^{\phantom{1}}\d_{\n\la}-\d_{\m\la}\d_{\n\eta}\right)\ .
\label{contraction}
\eeqn
It is trivial to check that 
$\Pi_0(x)\equiv\frac{1}{2^4}\langle{\cal O}_4^{\phantom{1}}(x)
{\cal O}_4(0)\rangle_{h=0}=(3N_i^{\ 2})(\pi^4x^8)$ \cite{GubserKleb}. 
To calculate $\Pi_1(x)$ first we expand
the symmetrized trace of ${\cal O}_8$ in terms of simple traces as
\eqn
\arr{rcl}
\displaystyle{{\cal O}_8(z)={\rm Str}_i\left[F^4-\frac{1}{4}(F^2)^2\right]}
&=&
\displaystyle{\frac{2}{3}{\rm tr}_i
\left[F_{\m\n}^{\phantom{\frac{1}{1}}}F^{\n\eta}F^{\m\la}F_{\la\eta}
+\frac{1}{2}F_{\m\n}F^{\n\eta}F_{\eta\la}F^{\la\m}\right.}\\
&&\ \ \ \ \ \ \displaystyle{\left.
-\frac{1}{4}F_{\m\n}F^{\m\n}F_{\eta\la}F^{\eta\la}
-\frac{1}{8}F_{\m\n}F_{\eta\la}F^{\m\n}F^{\eta\la}\right]}\ .
\earr
\label{Str}
\eeqn
Then using standard perturbative field theory techniques we have (see
also \cite{LiuTsey})
\eqn
\left\langle
{\cal O}_4^{\phantom{1}}(x){\cal O}_8(z){\cal O}_4(0)\right\rangle_{h=0}=
\frac{(3\times 2^8)N_i^{\ 3}}{\pi^8(x-z)^8z^8}\ ,
\label{3ptf}
\eeqn
and therefore 
\eqn
\Pi_1(x)=-\left(\frac{R_j}{2\D}\right)^4\frac{1}{2^7T_3}
\int d^4z\frac{(3\times 2^8)N_i^{\ 3}}{\pi^8(x-z)^8z^8}\ .
\eeqn
\begin{figure}
\centerline{\psfig{figure=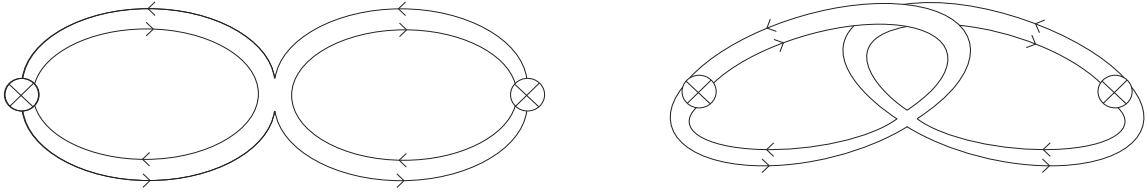,width=6in}}
\caption{\small{t'Hooft double line representation of the planar and 
non-planar Feynman diagrams associated with the three-point function
$\left\langle{\cal O}_4(x)
{\cal O}_8(z){\cal O}_4(0)\right\rangle_{h=0}$.}}
\label{fig1}
\end{figure} 
Notice that we have to be careful in applying Wick's theorem to 
obtain this result. The reason is that for large $N_i$ only contractions
between the fields in ${\cal O}_4$ and fields that are consecutive in the
trace expansion (\ref{Str}) of ${\cal O}_8$ will contribute. 
The other contractions
correspond to non-planar graphs and are subleading in the large 
$N_i$ limit\footnote{I thank Igor Klebanov for bringing this 
point to my attention.} (see figure). 
A consequence of this fact is that it was 
essential that we used the symmetrized trace in the effective action,
otherwise we would obtain a different answer. Using the result
\eqn
\int d^4u \frac{e^{ip\cdot u}}{u^8}=
\frac{\pi^2}{3\times 2^6}p^4\log{\left(p^2/\Lambda^2\right)}\ ,
\eeqn
a simple calculation gives the Fourier transform of $\Pi_1(x)$
\eqn
\Pi_1(p)=
-\left(\frac{R_j}{2\D}\right)^4\frac{N_i^{\ 3}}{(3\times2^{11})\pi^4}
p^8\left(\log{\left(p^2/\Lambda^2\right)}\right)^2\ ,
\label{Fourier}
\eeqn
where $\Lambda$ is a cut-off scale. Then the absorption cross section
is related by equation (\ref{disc}) to the momentum space two-point
function $\Pi(p)$. The result is 
\eqn
\s_i=\frac{\kappa^2\o^3N_i^{\ 2}}{32\pi}
\left[ 1-\left(\frac{R_j}{2\D}\right)^4
\frac{N_i\o^4}{(3\times2^3)T_3\pi^2}
\log{\left(\frac{\o}{\Lambda}\right)}\right]\ ,
\label{result}
\eeqn
which agrees exactly with the classical gravity 
prediction, including the numerical factors. 
We remark that in the perturbative field analysis it seems
natural to set the cut-off scale to $\Lambda=m_W$, while the strong coupled
gravity calculation suggests that
$\Lambda=\D/R_i^{\ 2}\sim m_W/\sqrt{gN}$. This
fact may be related to the existence of colour singlet condensates 
of W particles at strong coupling with a large binding energy 
\cite{GiddRoss,Gubser}. However, a better understanding of the cut-off
scale would require the extension of this calculation to the next order
\cite{GHKK}. 

In resume, we tested the AdS/CFT duality on the Coulomb branch by finding
agreement between the gravity and field theory absorption cross sections 
for the dilaton field near the $SU(N_i)$ IR conformal fixed points. This 
was done by determining the large $N$ (non-renormalized) $F^4$ terms in the
field theory IR effective action that arise from integrating the massive
W particles. Then the correction to the absorption cross section is 
related to a three-point function of chiral primary operators. To obtain
the correct result required large $N$ considerations as well as the use of
the symmetrized trace in the effective action. This result provides further
evidence for a non-renormalization theorem for three-point functions
of chiral primary operators [27-32].

\section*{Acknowledgments}

I would like to thank Lori Paniak and Igor Klebanov for many discussions. 
This work was supported by FCT (Portugal) under programme PRAXIS XXI 
and by the NSF grant PHY-9802484.

\end{document}